\documentclass[traditabstract]{aa}
\usepackage{amsmath, amssymb}% amsthm, amssymb}
\usepackage{txfonts}
\usepackage{graphicx}
\usepackage{color}
\usepackage{url}
\usepackage{natbib}

\newcommand{\rsun}{R_{\odot}}

\begin{document}

\title{The solar magnetic field since 1700.}
\subtitle{II. Physical reconstruction of total, polar and open flux}

\author{ J. {Jiang} \and R.~H. Cameron \and D. Schmitt  \and M. Sch\"{u}ssler}
\institute{Max-Planck-Institut f\"{u}r Sonnensystemforschung, 37191 Katlenburg-Lindau, Germany} 

\date{Received  / accepted }

\abstract{We have used semi-synthetic records of emerging sunspot groups based on
sunspot number data as input for a surface flux transport model to
reconstruct the evolution of the large-scale solar magnetic field and
the open heliospheric flux from the year 1700 onward. The statistical
properties of the semi-synthetic sunspot group records reflect those of
the observed the Royal Greenwich Observatory
photoheliographic results. These include correlations between the
sunspot numbers and sunspot group latitudes, longitudes, areas and tilt
angles. The reconstruction results for the total surface flux, the polar
field, and the heliospheric open flux (determined by a current sheet
source surface extrapolation) agree well with the available
observational or empirically derived data and reconstructions.  We
confirm a significant positive correlation between the polar field
during activity minimum periods and the strength of the subsequent
sunspot cycle, which has implications for flux transport dynamo models
for the solar cycle. Just prior to the Dalton minimum, at the end of the 
18th century, a long cycle was followed by a weak cycle.
We find that introducing a possibly `lost' cycle between 1793 and 1800 
leads to a shift of the minimum of the open flux by 15 years which is 
inconsistent with the cosmogenic isotope record.

}
\maketitle

\section{Introduction}
Information about the global properties of the solar magnetic field
during past centuries provides important constraints for models of the
Sun's global dynamo and is also relevant for studies of the past
terrestrial climate \citep{Usoskin08}.  However, regular synoptic
direct observations of the large-scale solar surface field and
space-based measurements of the interplanetary magnetic field cover only
a few decades.  The aim of our work is to reconstruct the Sun's
large-scale magnetic field and its heliospheric open flux from 1700
onwards on the basis of sunspot number records. The tool used for this
reconstruction is a surface flux transport model
\citep[SFTM,][]{Baumann04,Jiang10}. In contrast to
previous attempts at such reconstructions \citep[e.g.,][]{Wang05,
Schrijver02}, the source input for our SFTM model consists of
semi-synthetic records of sunspot groups whose properties (such as the
distributions of emergence latitudes, areas, and tilt angles) obey the
empirical statistical relationships with the activity cycle strength
studied in \citeauthor{Jiang10b} (\citeyear{Jiang10b} henceforth 
referred to as Paper~I).  The heliospheric open flux is extrapolated
from the distribution of surface flux using the current sheet source
surface (CSSS) model \citep{Zhao95a,Zhao95b}.  The obtained open flux can
be compared with reconstructions based upon the geomagnetic $aa$ index
\citep{Russell75, Lockwood99, Svalgaard05,Rouillard07} or upon
cosmogenic radionuclide data such as the concentration of $^{10}$Be in
ice cores \citep{Bard97, Caballero-Lopez04, McCracken07, Steinhilber10}.

This paper is organized as follows. The SFTM, the treatment of its
sources, and the CSSS extrapolation are described in Section~2. Observations and results from
\citet{Cameron10} (hereafter referred to as CJSS10) are used to validate the model in
Section~3. The results including the polar field, the total and the open
magnetic flux from 1700 onwards are presented in Section 4 and possible
errors are assessed in Section 5. Our conclusions are given in
Section~6.

\section{Model description}
\subsection{Solar surface flux transport model}
\label{sec:sft} 
The evolution of the large scale magnetic field on the solar surface can
be studied using two-dimensional flux transport models \citep{Devore84,
Wang89, Mackay00,Schrijver02, Baumann04}, which describe the passive
transport of the radial component of the magnetic field, $B$, under the
effect of differential rotation, $\Omega$, meridional flow, $v$
\citep{Babcock61}, and turbulent surface diffusivity, $\eta_H$
\citep{Leighton64}.  A slow decay due to the fact that diffusion occurs
in three dimensions \citep{Schrijver02} is modeled here in the manner
described in \cite{Baumann06}, introducing the parameter $\eta_r$.

The governing equation of the surface flux transport model (SFTM) is
\begin{eqnarray}
\label{eqn:SFT} \nonumber\frac{\partial B}{\partial t}=& &
-\Omega(\lambda,t)
                       \frac{\partial B}{\partial \phi}
         - \frac{1}{\rsun \cos\lambda}
              \frac{\partial}{\partial \lambda}[v(\lambda,t)
         B \cos \lambda] \\ \noalign{\vskip 2mm}
& & +\eta_{H} \left[\frac{1}{\rsun^2 \cos{\lambda}}
                \frac{\partial}{\partial \lambda}\left(\cos\lambda
          \frac{\partial B}{\partial \lambda}\right) +
     \frac{1}{\rsun^2 \cos^2{\lambda}}\frac{\partial^2 B}{\partial
     \phi^2}\right]\\ \noalign{\vskip 2mm} \nonumber
& & + S(\lambda,\phi,t) +D(\eta_r),
\end{eqnarray}
where $S(\lambda,\phi,t)$ is the source term of the magnetic flux, which
describes the emergence of bipolar magnetic regions as a function of
latitude $\lambda$, longitude $\phi$ and time $t$. $D(\eta_r)$ is a
linear operator describing the decay due to radial diffusion. The value
of $\eta_r$ is discussed in Section \ref{sec:calibration}. Concerning
the horizontal diffusivity, $\eta_H$, we use the reference value in
CJSS10 as $\eta_H=250$~km$^2$s$^{-1}$. The profiles of both differential
rotation, $\Omega$ \citep{Snodgrass83} and meridional flow, $v$
\citep{van_Ballegooijen98} are the same as in CJSS10: $\Omega=13.38-2.30
\sin^2 \lambda -1.62 \sin^4 \lambda$ (in degrees per day) and
\begin{equation}
v(\lambda)=\begin{cases}
11 \sin(2.4\lambda)~\mathrm{m~s^{-1}}& \text{for  $\vert\lambda\vert \le 75^{\circ}$},\\
   0 & \text{otherwise}.
\end{cases}
\label{eqn:mer}
\end{equation}
While we take the meridional flow velocity to be
time-independent, latitudinal inflows towards the active regions are
taken into account via a reduction of the tilt angle as described in the
next subsection \citep[see also CJSS10; ][]{Jiang10c,Cameron10b}. 
For the initial field distribution, we follow \citet{van_Ballegooijen98} and
CJSS10. The strength of the initial field is determined by the parameter
$B_0$ corresponding to the field strength at the poles. The values used
are discussed in Section \ref{sec:results}.

The solution of the linear Equation (\ref{eqn:SFT}) may be expressed in
terms of spherical harmonics
\begin{equation}
B(\lambda, \phi,
t)=\sum_{l=1}^{\infty}\sum_{m=-l}^{l}a_{lm}(t)Y_l^{m}(\lambda, \phi).
\label{eqn:sphe}
\end{equation}
The axial dipole ($l=1, m=0$) and equatorial dipole components ($l=1,
m=\pm 1$) are considered in the presentation of our results. We use the same
definitions as \cite{Wang05} to calculate the solar surface axial
($D_{\textrm{ax}}$) and equatorial ($D_{\textrm{eq}}$) dipole
strengths. The polar field is defined as the average over a polar cap of
$15^{\circ}$ latitude extension. The total surface flux is obtained by
integrating the unsigned magnetic field over the whole surface.

\subsection{Sources of magnetic flux}
\label{sec:SG_emergence} 
\begin{figure}
\centering
\resizebox{\hsize}{!}{\includegraphics[angle=0]{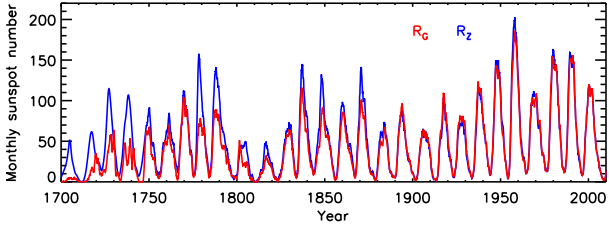}}
\caption{12-month running average of monthly Wolf ($R_Z$, blue) and
group ($R_G$, red) sunspot numbers during 1700--2010. $R_G$
in cycle 23 uses the values of $R_Z$.}
\label{fig:WolfVSgroup}       % Give a unique label
\end{figure}

Here we describe how we model the magnetic flux source term,
$S(\lambda,\phi,t)$, in Equation (\ref{eqn:SFT}). We use the
semi-synthetic sunspot group records determined in Paper~I on the basis
of either the group sunspot number, $R_G$ \citep{Hoyt98} or the Wolf
sunspot number, $R_Z$, the 12-month running averages of which are shown
in Figure \ref{fig:WolfVSgroup} from 1700 onwards. Each sunspot group in
the semi-synthetic record is taken to represent a bipolar
magnetic region (BMR).  The area of a BMR, being the sum of the umbral,
penumbral and facular areas, is calculated from the area of the
corresponding sunspot group using the empirical relation found by
\citet{Chapman97}.  The tilt angle of the BMR, $\alpha$, is assumed to
be 70\% of that of the sunspot group. This factor (70\%) was empirically
determined in CJSS10 and reflects the effect of the localized inflows
into active regions.

In Paper I, the tendency of sunspots to appear in activity nests was
measured by considering which combination of uniformly distributed
random emergence longitudes and completely ordered longitudes match the
degree of non-randomness in the sunspot record of the Royal Greenwich
Observatory since 1874. Here we follow the same approach and use both
the semi-synthetic datasets with ordered and random
emergence longitudes. How the resulting fields are
combined is discussed in Section 3.

The magnetic field distribution of each new BMR is given by 
$B(\lambda,\phi)=B^+(\lambda,\phi)-B^-(\lambda,\phi)$ with
\begin{eqnarray}
B^{\pm}(\lambda,\phi)=B_{\mathrm{max}} \left(\frac{0.4
\Delta\beta}{\delta}\right)^2
  \mathrm{\exp}\{-\frac{2[1-\cos\beta_{\pm}(\lambda,\phi)]}{\delta^2}\},
\end{eqnarray}
where $\beta_{\pm}(\lambda,\phi)$ are the heliocentric angles between
the location of the sunspot group,$(\lambda,\phi)$, and the centers of
each polarity, $(\lambda_{\pm},\phi_{\pm})$.  $\Delta
\beta$ is the separation between the two polarities. We take
$\Delta\beta=0.45A_R^{1/2}$, where $A_R$ is the total area of the
sunspot groups, and $\delta=4^\circ$ \citep{van_Ballegooijen98,
Baumann04}. $B_\mathrm{max}$ is the peak field strength of a
BMR. Following CJSS10, we take $B_\mathrm{max}=374$~G.

$\Delta\beta$ and the tilt angle, $\alpha$, determine the latitudinal
separation, $\Delta\beta\sin\alpha$, of the two polarities of a BMR.
This quantity is important for the axial dipole moment of the global
magnetic field and affects the polar field and the open flux during
activity minima. The longitudinal separation, given by
$\Delta\beta\cos\alpha(\cos\lambda)^{-1}$, dominates the equatorial
dipole moment and the open flux during activity maximum phases.

Associated with the $n^{\mathrm{th}}$ BMR in the semi-synthetic 
sunspot record is the time $t_n$ when the BMR appears. The contribution to the
source term $S(\lambda,\phi,t)$ from the $n^{\mathrm{th}}$ BMR is taken to be
$\delta(t - t_n) \left(B^{+}(\lambda,\phi)-B^{-}(\lambda,\phi)\right)$, where $\delta$ is
the Dirac delta function. This corresponds to modeling the sunspot group 
emergence process as being instantaneous.
 
\subsection{Extrapolation into the heliosphere}
The SFTM describes the evolution of the magnetic field on the
Sun's surface.  To obtain the heliospheric open flux we have to
extrapolate the surface field outward. Since the often used potential
field source surface model does not well represent the Ulysses
spacecraft data \citep{Schuessler06}, we use the better suited current
sheet source surface (CSSS) extrapolation \citep{Zhao95a,Zhao95b}.

There are three parameters in the CSSS extrapolation, namely, one
associated with the thickness of the current sheet, $a$, the radius of the cusp
surface, $R_{\rm{cs}}$, and the radius of the source surface,
$R_{\rm{ss}}$. The values for the three parameters are taken from the
reference case in CJSS10, i.e., $a=0.2\rsun$, $R_{\rm{cs}}=1.55\rsun$,
$R_{\rm{ss}}=10.0\rsun$. The location of the cusp surface determines the
amount of the open flux, which is calculated by the integration of
unsigned magnetic field over the whole cusp surface.

\section{Comparison of the model for the time period from 1913 to 1986}
\label{sec:calibration} 
In CJSS10, we used the SFTM with a source term based on the actually
observed sunspot longitudes, latitudes, areas and cycle-averaged tilt
angles to reconstruct the surface field and open flux for the period
1913--1986.  The time evolution of the open flux derived from
geomagnetic indices \citep{Lockwood03, Lockwood09b}, and the reversal
times of polar fields \citep{Makarov03} were well reproduced.
We therefore begin by comparing the results obtained using the current model 
based on sunspot numbers with those in CJSS10.

\begin{figure}
\centering
\resizebox{\hsize}{!}{\includegraphics{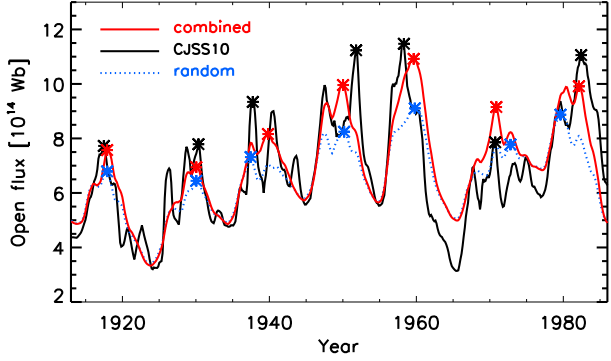}}
\caption{Open flux from CJSS10 (solid black curve), from the semi-synthetic
sunspot group record with random longitudes (dotted blue curve),
and from a combination of semi-synthetic models with random and ordered 
longitudes 
$B_{\textrm{com}}=(1-\mathrm{c})B_{\textrm{ran}}+\mathrm{c}B_{\textrm{ord}}$ with
$\mathrm{c}=0.15$ (red curve). The asterisks show the maximum value for 
each cylce. Results using the semi-synthetic data are averages over
20 realizations.}
\label{fig:calibrate_c}       % Give a unique label
\end{figure}

To do so we have to consider the tendency for sunspots to occur in activity
nests at specific longitudes, because this affects equatorial dipole
moment and the open flux around activity maxima.  We performed
simulations with BMRs emerging at randomly distributed longitudes and
simulations where all BMRs appear at longitudes 90$^{\circ}$ in the
northern hemisphere and 270$^{\circ}$ in the southern hemisphere. Since
both the SFTM and the CSSS extrapolation are linear, we can combine the
fields from the two kinds of simulations using
\begin{equation}
B_{\textrm{com}}(R_{\textrm{cs}}, \theta, \phi,
t)=(1-\mathrm{c})B_{\textrm{ran}}(R_{\textrm{cs}},\theta, \phi, t)
+\mathrm{c}B_{\textrm{ord}}(R_{\textrm{cs}},\theta,\phi, t).
\end{equation}
where $\mathrm{c}$ is a measure of the strength of the activity nesting.  In
Paper I we found that the value $\mathrm{c}=0.15$ is appropriate for a 
determination of the open flux. This is confirmed here.  Figure
\ref{fig:calibrate_c} shows the evolution of the open flux from CJSS10
compared with cases with $\mathrm{c}=0$ (purely random) and $\mathrm{c}=0.15$. The case
with random longitudes generates too low open flux at solar maxima. 
The rms difference between the results from CJSS10 is 
$1.19 \times 10^{14}$~Wb when $c=0$ (the purely random case) 
compared to  $1.09 \times 10^{14}$~Wb when  $\mathrm{c}=0.15$. 
When we consider only the cycle maximum values of the open flux for each cycle, 
the difference is greater, with the rms deviation falling from 
$1.93 \times 10^{14}$~Wb for $c=0$ to $1.00 \times 10^{14}$~Wb for $c=0.15$.

\begin{figure*}
\resizebox{\hsize}{!}{\includegraphics{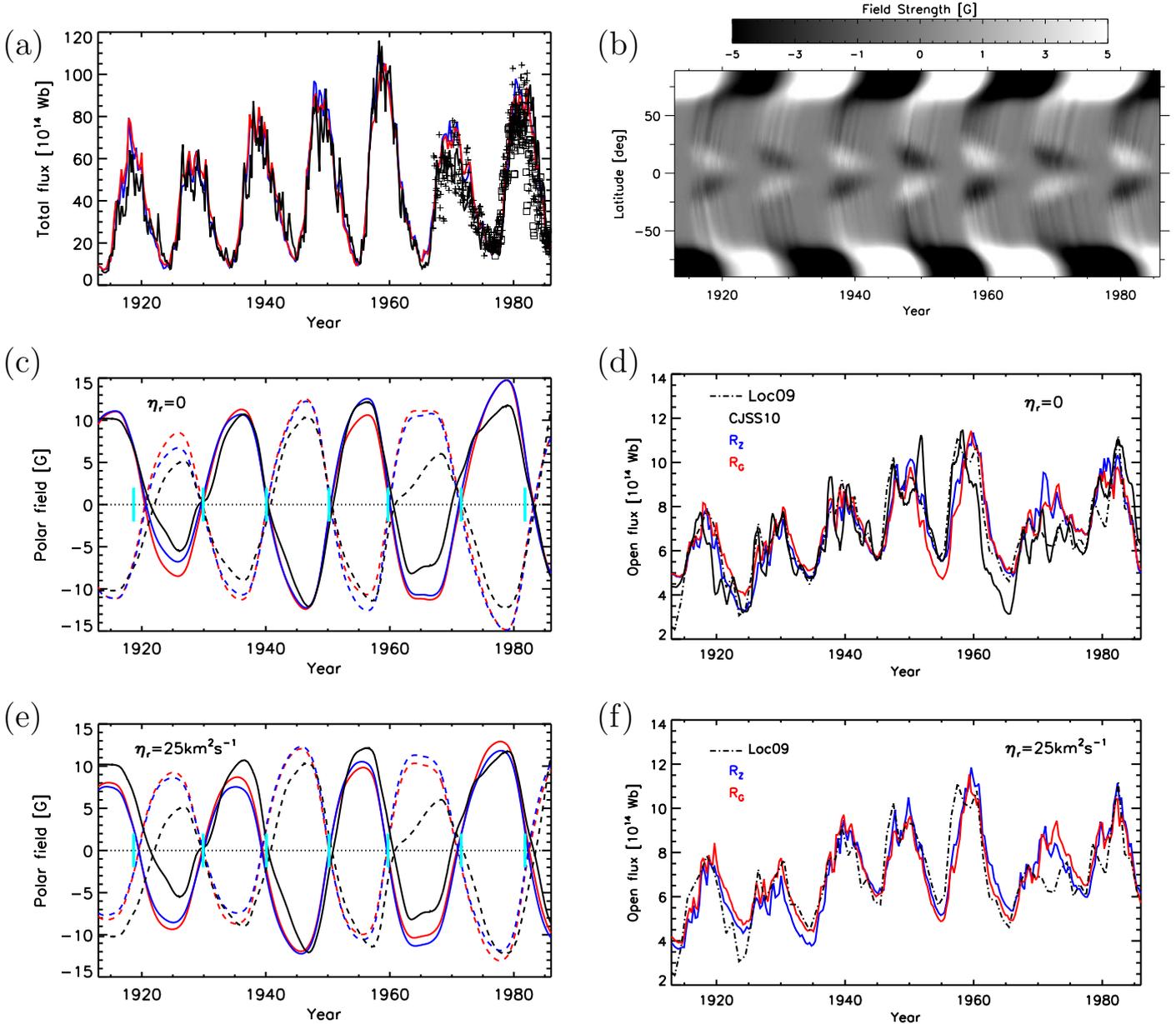}}
\caption{Comparison of the model results based on $R_Z$ (blue curves)
and $R_G$ (red curves) with the reconstruction of CJSS10 (black curves)
for the period 1913--1986. Panels (a)--(d) show the case without radial
diffusion. (a) Total surface flux. Square and plus symbols during the
last 2 cycles denote the observations from the Wilcox and Mount Wilson
solar observatories, respectively. (b) Time-latitude diagram of the
longitudinally averaged signed surface magnetic field from the model
based on $R_Z$. (c) Polar field. Dashed and solid curves are for the
southern and northern hemispheres, respectively. The cyan vertical lines
indicate the inferred times of polar field reversals from
\citet{Makarov03}. (d) Open heliospheric flux. The dash-dotted curve
represents the result inferred from the geomagnetic $aa$ index
\citep{Lockwood09b}. (e)--(f): results with radial diffusivity
$\eta_r=25$~km$^2$s$^{-1}$. (e) Polar field. (f) Open heliospheric
flux.}
\label{fig:1913_1986}       % Give a unique label
\end{figure*}

Next we compare our simulations during 1913 -- 1986 (cycles 12--21)
based on $R_G$ and $R_Z$ with the results from CJSS10 and with the
observed data. Note that all
results shown are based on averages over 20 sets of random realizations
of the semi-synthetic sunspot records.  We first set the radial
diffusivity to $\eta_r=0$. Figure \ref{fig:1913_1986}(a) compares the
time evolution of total flux from CJSS10 with the models based on $R_G$
and $R_Z$, respectively. The three curves almost overlap. For the period
from the early 1970s onwards, the direct magnetic measurements from the
Mount Wilson and Wilcox observatories are also shown.  Figure
\ref{fig:1913_1986}(c) shows the evolution of the polar field for both
hemispheres.  Since the input of the BMRs emergences is randomly
distributed on the two hemispheres, our reconstructed north and south
polar fields are similar. The reversal times (indicated by cyan vertical
lines) are similar to those given by \citet{Makarov03}. The largest
differences between the models occur in cycle 19.  Figure
\ref{fig:1913_1986}(d) shows the evolution of the modeled open flux in
comparison with that inferred from the geomagnetic $aa$ index
\citep{Lockwood09b}. For the whole time period, the rms difference
between the inferred values from the $aa$ index and CJSS10, $R_Z$ and
$R_G$ are 0.99, 0.92, 0.96 $\times 10^{14}$~Wb, respectively,
corresponding to about 14\% of the average value of 6.92
$\times10^{14}$~Wb of the \citet{Lockwood09b} data.

The time-latitude plot of the longitudinal averaged signed photospheric
field (magnetic butterfly diagram) during this time period is shown in
Figure \ref{fig:1913_1986}(b) for the reconstruction based on $R_Z$. As
found in previous studies, the latitude separation of the two polarities
leads to a net flux when azimuthally averaged. The advection of this
flux to the poles reverses the polar fields each cycle.  The regular
polar field reversals and the anti-phase between the polar field and the
low latitude field are well reproduced.

These results show that our SFTM with sources based on the sunspot
numbers $R_Z$ or $R_G$ is consistent with the reconstruction of CJSS10,
which used the recorded properties of the actually sunspot
groups, and also compares well with the observed photospheric and the
heliospheric magnetic field.

Because the sunspot number data are less reliable before 1849
\citep{Vaquero07, Svalgaard10} we have chosen to use a non-vanishing
value for the radial diffusivity $\eta_r$. This was found to be
necessary because when $\eta_r=0$ is used, the e-folding decay time 
due to $\eta_{H}$ alone is approximately 4000 years. This timescale
was obtained numerically, and is considerably longer than the simple estimate 
$\pi R_{\sun}^2/\eta$ because the meridional flow tends to keep
the magnetic field at the two poles separated.
The long e-folding time means that the field at any given time is affected by errors in the 
sunspot numbers at all previous times.  This is very undesirable because noisy 
data at the beginning of the dataset then contribute for the entire period covered
by the simulations without significant damping.  Introducing a weak
radial diffusivity $\eta_r=25$~km$^2$s$^{-1}$ (which leads to a decay
time about 20 yr, as found by \citeauthor{Baumann06}, \citeyear{Baumann06}) 
is aimed at keeping $\eta_r$ small while still being
able to sensibly use the early data. Figure \ref{fig:1913_1986}(e) shows
the polar field evolution with the weak radial diffusion
included. Compared to Figure \ref{fig:1913_1986}(c), the largest
differences are on the order of 20\%. The time evolution of the
corresponding open fluxes is shown in Figure
\ref{fig:1913_1986}(f). During the minimum phases of some cycles, the
model is now closer to the data inferred from the $aa$ index (e.g. $R_G$
model around 1955 and $R_Z$ model around 1976), while other deviate more
strongly (e.g., $R_Z$ model around 1924 and 1936). For the whole time
period, the rms deviations for the models based on $R_Z$ and $R_G$ are
0.93 and 0.96 in $10^{14}$~Wb, respectively.

%\ifnum{2<1}
CJSS10 found a strong correlation between the polar field of cycle $n$
and the strength of the subsequent cycle $n+1$. Without radial
diffusion, the correlation coefficients are 0.80 and 0.52 for the $R_Z$
and $R_G$, cases respectively.  In our case with a sample size of 7, a
significance level of $p=$~0.05 corresponds to $r=$~0.74. For all cases,
there is no correlation between the polar field around the activity
minimum of cycle $n$ and the strength of the same cycle. The
introduction of radial diffusivity and the corresponding decay of the
polar field decreases the correlation between the polar field and its
subsequent cycle strength and somewhat increases the correlation between
the polar field of cycle $n$ and the strength of the same cycle. All
these correlation coefficients do not exceed the level required for
significance at the $p=0.05$ level.
%\fi

We conclude that the SFTM with input data based on sunspot numbers
effectively describes the solar magnetic field since 1913. Although
the introduction of a non-vanishing radial diffusivity leads to
some possibly undesired decay of the polar field, it decreases the error
caused by the noisy early sunspot numbers.  The photospheric
and the heliospheric field are reasonably well reproduced.  In the
next section, we present the results back to 1700.

\section{Results}
\label{sec:results}
\subsection{Time evolution of the reconstructed field}
\begin{figure}
\resizebox{\hsize}{!}{\includegraphics[width=0.7\textwidth]{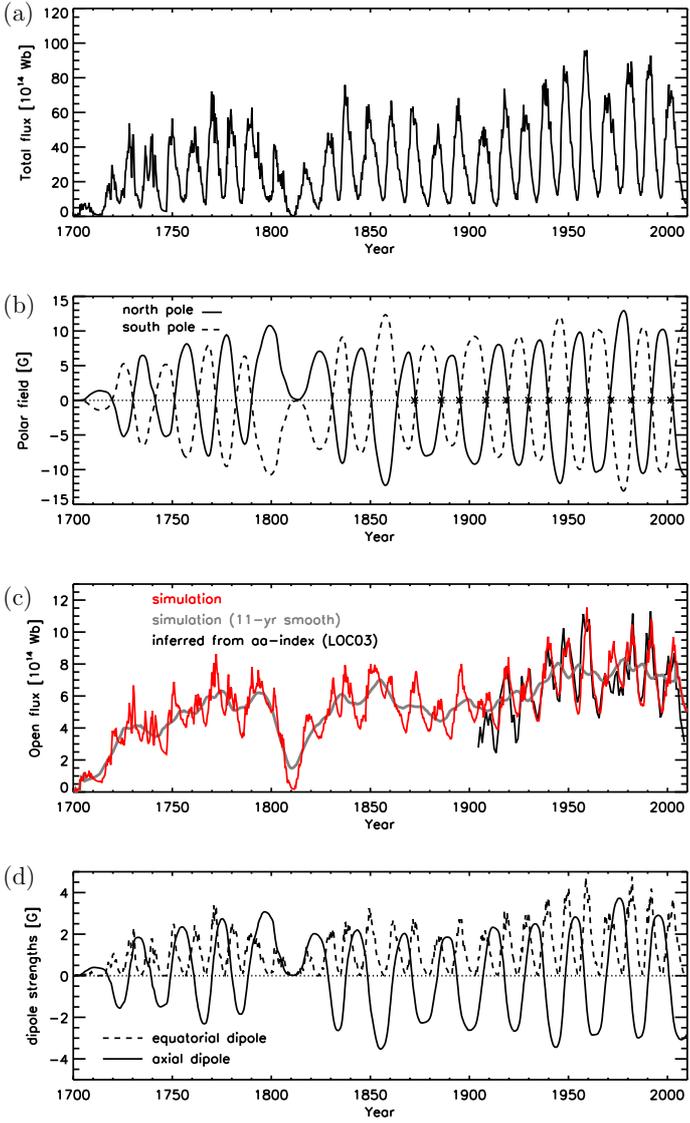}}
\caption{Reconstruction based on $R_G$ during 1700--2010. (a) Total
surface flux. (b) Polar field. Solid and dashed curves are for the
northern and southern hemisphere, respectively. (c) Open heliospheric
flux. The red curve gives the reconstruction, the grey curve is the
11-yr running average. The open flux as inferred from the geomagnetic
$aa$ index \citep{Lockwood09b} is represented by the black
curve. (d) Axial (solid curve) and equatorial (dashed curve) dipole
field strength.}
\label{fig:group_1700_2010}       % Give a unique label
\end{figure}

\begin{figure}
\resizebox{\hsize}{!}{\includegraphics{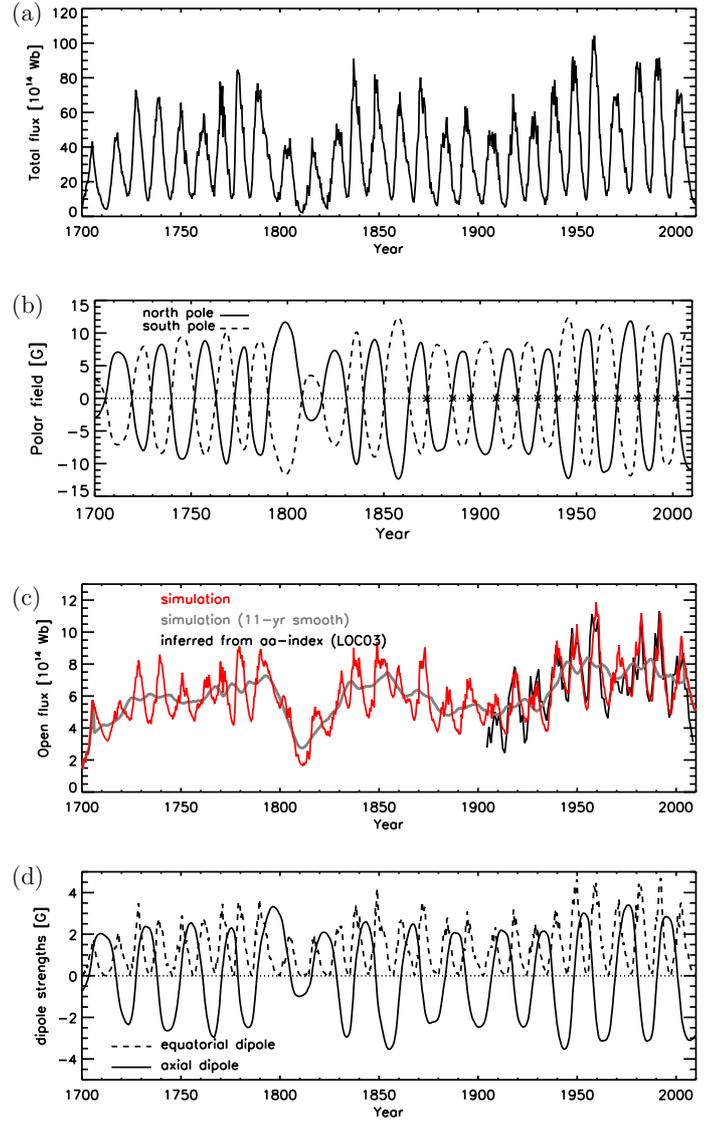}}
\caption{Same as Figure \ref{fig:group_1700_2010}, but for the
reconstruction based on $R_Z$.}
\label{fig:wolf_1700_2010}       % Give a unique label
\end{figure}

Figures \ref{fig:group_1700_2010} and \ref{fig:wolf_1700_2010} show time
series of various properties of the reconstructed solar magnetic field,
based on $R_G$ and $R_Z$, respectively. Yearly values for the 
reconstructed open fluxes are given in Tables~\ref{table:1} and \ref{table:2} 
(electronic version only).   
All results shown correspond
to averages over 20 sets of random realizations of the semi-synthetic
sunspot records.  The evolution of the total unsigned flux (panels a),
polar field (panels b), open flux (panels c), axial and equatorial
dipole field strength (panels d) are shown.  The first $\sim20$ yrs of these
series are still affected by the choice of the amplitude of the initial
field, $B_0$.  Since cycle $-4$ (1700--1712) is very weak
in the $R_G$ data, we set $B_0=0$ in this case. For
$R_Z$ we used $B_0=3$~$G$.


\onltab{1}{
\begin{table}
\caption{Reconstructed open flux based on $R_{\mathrm G}$}
\label{table:1}
\centering 
\begin{tabular}{c r r r r r r r r r r}
\hline\hline
Year &  \\
\hline
1700--1709& 0.01& 0.08& 0.16& 0.72& 0.96& 0.95& 1.20& 1.05& 0.98& 0.94\\
1710--1719& 0.88& 0.76& 0.71& 0.69& 0.74& 0.97& 1.32& 2.29& 1.80& 3.42\\
1720--1729& 3.46& 4.00& 3.29& 3.11& 3.31& 2.88& 4.09& 4.56& 4.16& 3.86\\
1730--1739& 6.12& 4.88& 4.21& 3.57& 3.34& 3.55& 4.71& 3.55& 3.04& 4.82\\
1740--1749& 3.33& 4.51& 3.56& 3.30& 2.69& 2.53& 2.43& 3.21& 6.01& 5.37\\
1750--1759& 5.43& 5.81& 5.12& 5.04& 4.32& 3.99& 3.93& 4.26& 4.54& 5.40\\
1760--1769& 5.07& 5.47& 5.48& 5.46& 5.24& 4.83& 4.29& 4.30& 5.06& 5.65\\
1770--1779& 8.06& 8.07& 9.96& 6.59& 6.74& 5.43& 5.13& 5.65& 6.29& 6.57\\
1780--1789& 5.01& 5.39& 5.09& 3.92& 3.31& 3.20& 3.95& 4.77& 6.08& 5.99\\
1790--1799& 6.99& 7.18& 7.21& 7.68& 7.04& 6.57& 5.92& 5.50& 5.61& 5.20\\
1800--1809& 5.17& 6.35& 5.54& 3.68& 3.29& 3.01& 3.15& 1.46& 0.72& 0.28\\
1810--1819& 0.13& 0.13& 0.44& 1.27& 1.94& 2.59& 3.95& 4.15& 4.23& 4.62\\
1820--1829& 4.43& 4.00& 3.67& 3.42& 3.59& 3.73& 3.93& 3.93& 5.23& 4.84\\
1830--1839& 6.68& 7.10& 5.74& 5.07& 4.61& 5.23& 6.84& 7.31& 6.77& 5.66\\
1840--1849& 5.75& 5.11& 4.37& 4.17& 3.86& 3.86& 5.10& 4.49& 6.89& 7.32\\
1850--1859& 8.70& 7.32& 7.63& 7.66& 7.18& 6.69& 6.35& 6.32& 7.46& 7.08\\
1860--1869& 7.04& 6.59& 6.21& 5.67& 4.59& 4.07& 3.83& 3.83& 4.14& 5.55\\
1870--1879& 6.61& 5.96& 7.80& 8.09& 6.48& 6.08& 4.64& 4.52& 4.42& 4.11\\
1880--1889& 4.21& 5.06& 5.74& 5.44& 5.32& 5.46& 5.51& 4.17& 3.62& 3.49\\
1890--1899& 3.25& 4.21& 4.80& 5.53& 6.71& 7.15& 7.31& 7.13& 6.08& 5.43\\
1900--1909& 4.88& 4.71& 4.41& 4.50& 5.42& 5.66& 6.49& 6.21& 6.51& 6.17\\
1910--1919& 6.26& 5.35& 4.41& 4.02& 3.91& 5.20& 6.34& 7.07& 6.28& 8.24\\
1920--1929& 7.21& 6.23& 5.82& 5.22& 4.91& 5.35& 6.89& 6.20& 7.61& 6.27\\
1930--1939& 6.19& 5.43& 4.80& 4.58& 4.55& 4.67& 5.69& 8.38& 8.10& 9.69\\
1940--1949&10.70& 8.99& 7.35& 6.68& 6.18& 6.17& 7.68& 9.42& 8.61& 7.85\\
1950--1959&10.50& 7.63& 6.43& 5.97& 5.19& 5.24& 6.87& 8.39& 9.42&10.31\\
1960--1969& 9.77& 8.59& 7.06& 5.92& 5.42& 5.14& 5.16& 6.37& 6.66& 7.29\\
1970--1979& 7.87& 7.48& 8.31& 9.03& 7.91& 6.71& 6.37& 6.28& 7.30& 8.82\\
1980--1989& 9.81& 8.14&11.16& 9.30& 8.99& 6.79& 6.09& 6.17& 6.44& 8.30\\
1990--1999& 8.78& 8.92& 9.14& 8.12& 6.34& 5.83& 4.95& 4.71& 5.74& 6.17\\
2000--2009& 7.38& 7.89& 8.31& 8.30& 7.90& 6.73& 6.89& 6.12& 5.51& 5.22\\
\end{tabular}
\end{table}
}

\onltab{2}{
\begin{table}
\caption{Reconstructed open flux based on $R_{\mathrm Z}$}
\label{table:2}
\centering 
\begin{tabular}{c r r r r r r r r r r}
\hline\hline
Year &  \\
\hline
1700--1709& 1.79& 1.94& 2.59& 2.98& 4.30& 4.96& 5.42& 4.53& 4.24& 4.64\\
1710--1719& 3.87& 3.47& 3.30& 3.28& 3.33& 4.22& 4.52& 3.99& 5.34& 4.89\\
1720--1729& 5.19& 4.99& 5.18& 4.55& 4.45& 4.93& 6.86& 6.36& 8.01& 7.55\\
1730--1739& 6.85& 5.84& 5.14& 4.76& 4.58& 4.90& 5.80& 6.80& 6.32& 8.23\\
1740--1749& 7.70& 6.61& 5.77& 4.60& 4.66& 4.32& 4.56& 5.05& 5.81& 5.08\\
1750--1759& 6.30& 5.62& 6.34& 6.25& 5.17& 4.82& 5.00& 4.95& 5.01& 4.95\\
1760--1769& 4.80& 6.20& 6.74& 5.86& 7.19& 5.62& 5.37& 5.31& 6.47& 7.54\\
1770--1779& 7.34& 6.65& 6.02& 6.84& 5.24& 4.46& 4.40& 6.21& 8.05& 9.64\\
1780--1789& 8.58& 7.95& 7.23& 5.95& 5.24& 4.84& 5.96& 6.51& 6.96& 8.51\\
1790--1799& 9.82& 6.94& 7.30& 7.71& 6.97& 6.92& 6.50& 5.92& 5.66& 5.33\\
1800--1809& 5.58& 6.14& 5.69& 5.30& 4.95& 4.90& 4.54& 2.95& 2.48& 2.07\\
1810--1819& 1.81& 1.76& 1.93& 2.34& 2.11& 3.23& 4.74& 4.86& 4.66& 5.28\\
1820--1829& 4.91& 3.98& 3.69& 3.51& 3.63& 3.73& 4.49& 4.88& 4.67& 5.84\\
1830--1839& 7.83& 7.63& 6.16& 5.77& 5.43& 5.82& 7.17& 8.72& 8.26& 7.49\\
1840--1849& 8.57& 6.41& 5.51& 4.93& 4.48& 5.11& 5.17& 6.06& 8.37&11.06\\
1850--1859& 9.33& 9.64& 8.72& 8.37& 7.98& 6.97& 6.36& 6.26& 7.20& 7.56\\
1860--1869& 7.22& 6.04& 5.55& 5.38& 5.67& 5.53& 4.69& 4.39& 4.63& 5.92\\
1870--1879& 6.62& 8.21& 9.37& 9.30& 7.73& 5.45& 4.74& 4.83& 4.23& 3.94\\
1880--1889& 4.44& 5.38& 5.56& 6.19& 5.42& 5.31& 5.24& 5.06& 4.18& 3.85\\
1890--1899& 3.71& 4.28& 5.64& 5.83& 5.46& 5.66& 7.11& 5.46& 5.85& 4.83\\
1900--1909& 4.62& 4.10& 4.00& 4.09& 4.84& 5.27& 4.78& 4.28& 4.39& 5.66\\
1910--1919& 5.55& 4.89& 4.36& 4.03& 3.92& 5.34& 4.60& 6.75& 6.79& 6.16\\
1920--1929& 6.53& 5.33& 4.96& 4.45& 4.38& 5.06& 5.83& 6.52& 5.70& 5.44\\
1930--1939& 6.53& 4.96& 4.59& 4.27& 4.15& 4.70& 5.37& 8.08& 7.62& 8.32\\
1940--1949& 8.78& 8.95& 8.31& 7.17& 6.65& 6.31& 7.22&10.75& 9.25& 9.58\\
1950--1959& 9.36& 9.05& 8.03& 5.93& 5.30& 5.46& 7.30& 8.83&10.47&10.95\\
1960--1969&10.78& 9.63& 7.63& 6.88& 6.10& 5.63& 5.55& 6.75& 7.94& 6.11\\
1970--1979& 8.04& 8.05& 8.31& 8.30& 7.47& 6.85& 6.38& 6.18& 7.22& 8.21\\
1980--1989&10.27& 8.82&10.72& 9.00& 7.92& 6.55& 5.95& 5.87& 6.44&10.18\\
1990--1999& 9.07&10.13& 9.40& 7.89& 6.71& 5.55& 5.00& 5.05& 5.79& 5.07\\
2000--2009& 7.54& 6.96&10.36& 8.11& 7.19& 6.77& 6.60& 5.68& 5.37& 5.06\\
\end{tabular}
\end{table}
}

The polar field displays regular reversals, except for the $R_G$ case
during the Dalton minimum, when a reversal appears to fail.  The flux
transported to the poles during the weak cycle 5 just cancels the strong
polar field generated by the strong and long cycle 4.  On the other
hand, a clear reversal occurs in the case based on $R_Z$
(cf. Figure~\ref{fig:wolf_1700_2010}). Owing to the uncertainties in the
sunspot number record, it cannot be determined with any confidence
whether the polar field actually reversed or not. 
The grey curves in Panels (c) of Figures \ref{fig:group_1700_2010} and
\ref{fig:wolf_1700_2010} are the 11-yr running average of the open
flux. The long-term trend is compared with other independent
reconstructions in Section \ref{sec:compare_field}.  

Panels (d) of Figures \ref{fig:group_1700_2010} and
\ref{fig:wolf_1700_2010} show the time evolution of the axial dipole
field, $D_{\textrm{ax}}$, and the equatorial dipole field,
$D_{\textrm{eq}}$.  $D_{\textrm{eq}}$ is approximately in phase
with the evolution of the sunspot number and the total flux. Since we
define the polar field as the average over latitudes poleward of
$\pm75^{\circ}$, $D_{\textrm{ax}}$ varies in phase with the polar
field. Due to the persistent emergence of sunspot groups in different
locations, the evolution of $D_{\textrm{eq}}$ is noisier than that of
$D_{\textrm{ax}}$. This is partly mitigated by the fact that we show
the average over 20 semi-synthetic sunspot group record realizations.

\subsection{Correlations with sunspot numbers}

\begin{figure*}
\resizebox{\hsize}{!}{\includegraphics{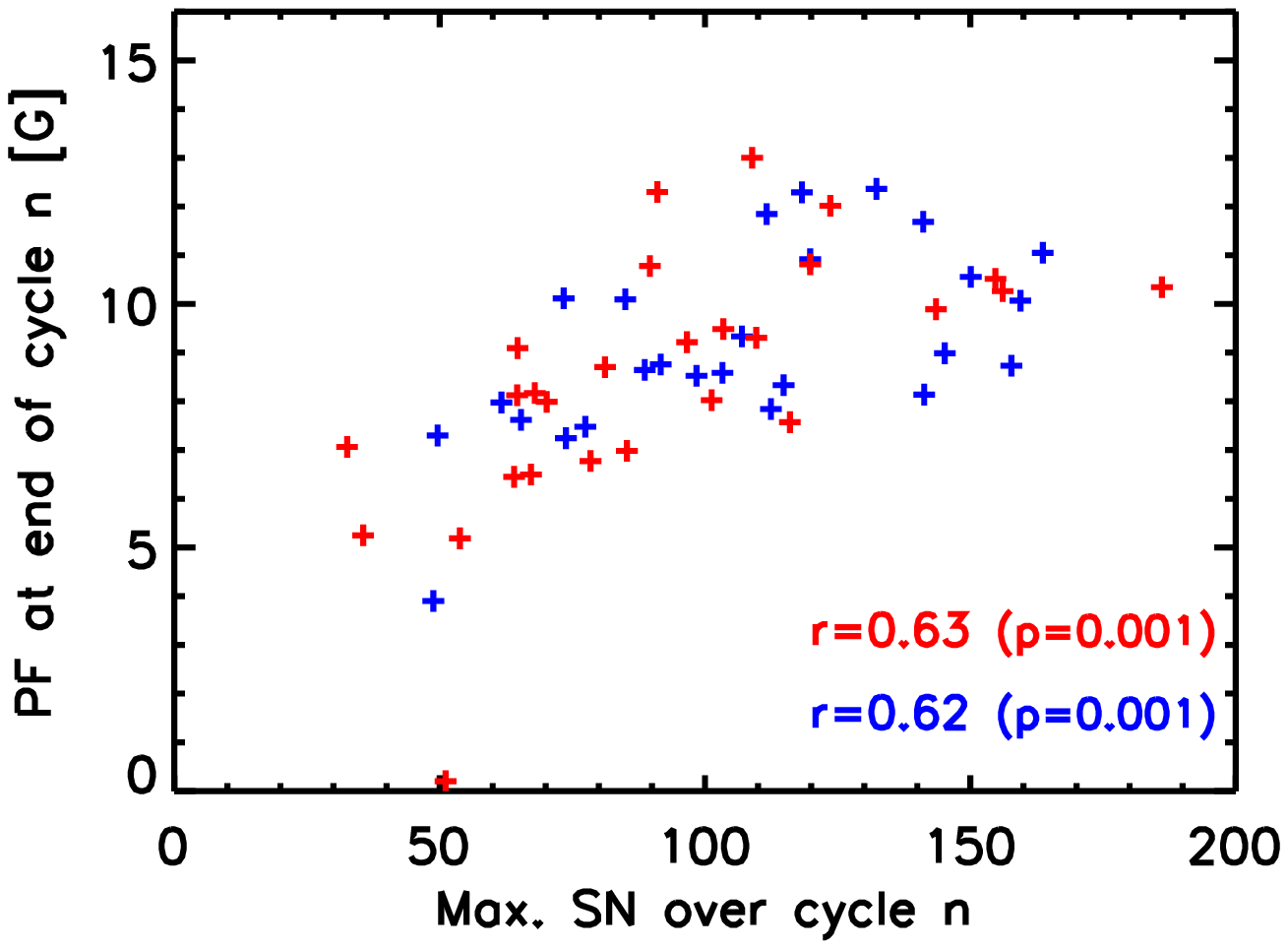}\includegraphics{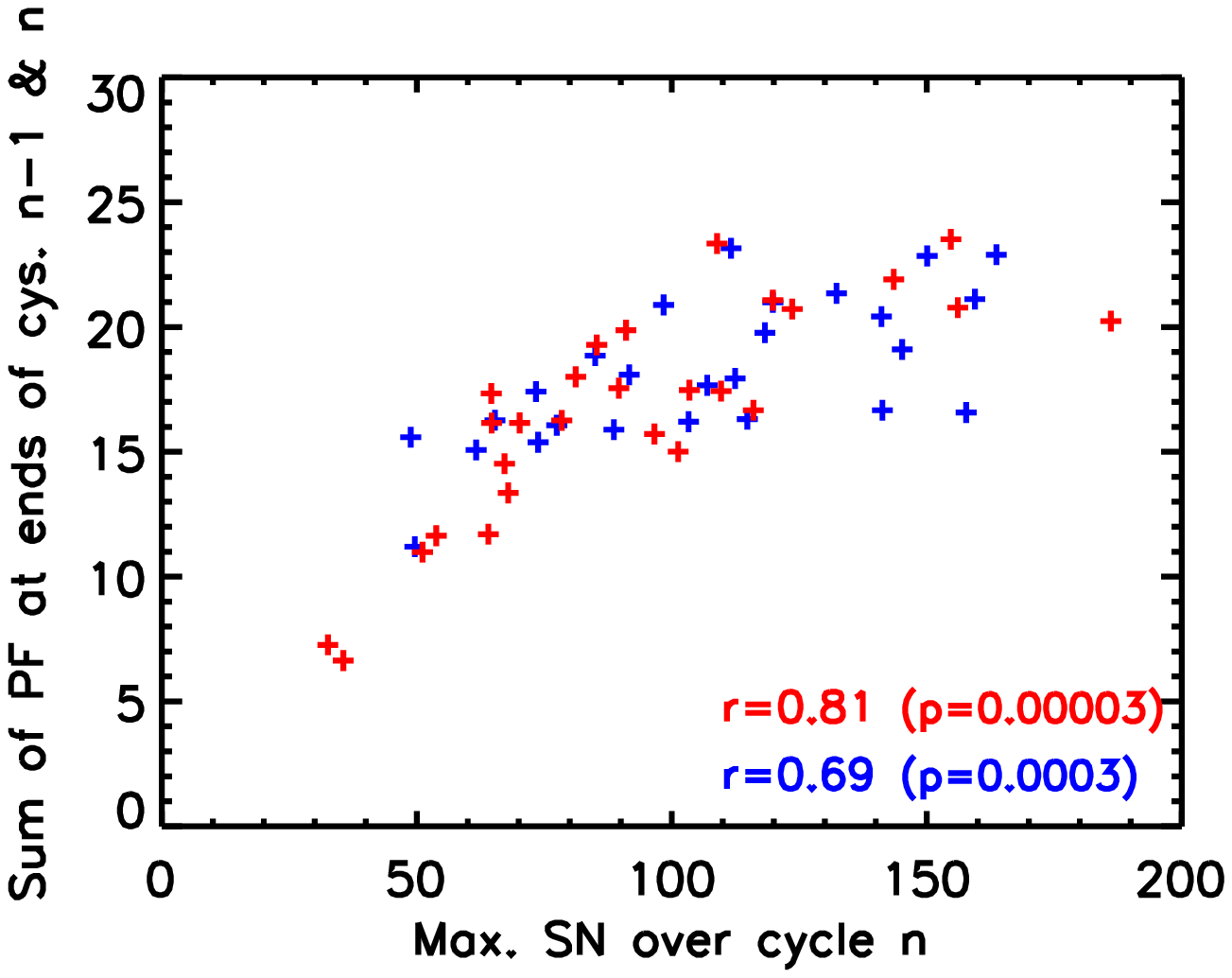}\includegraphics{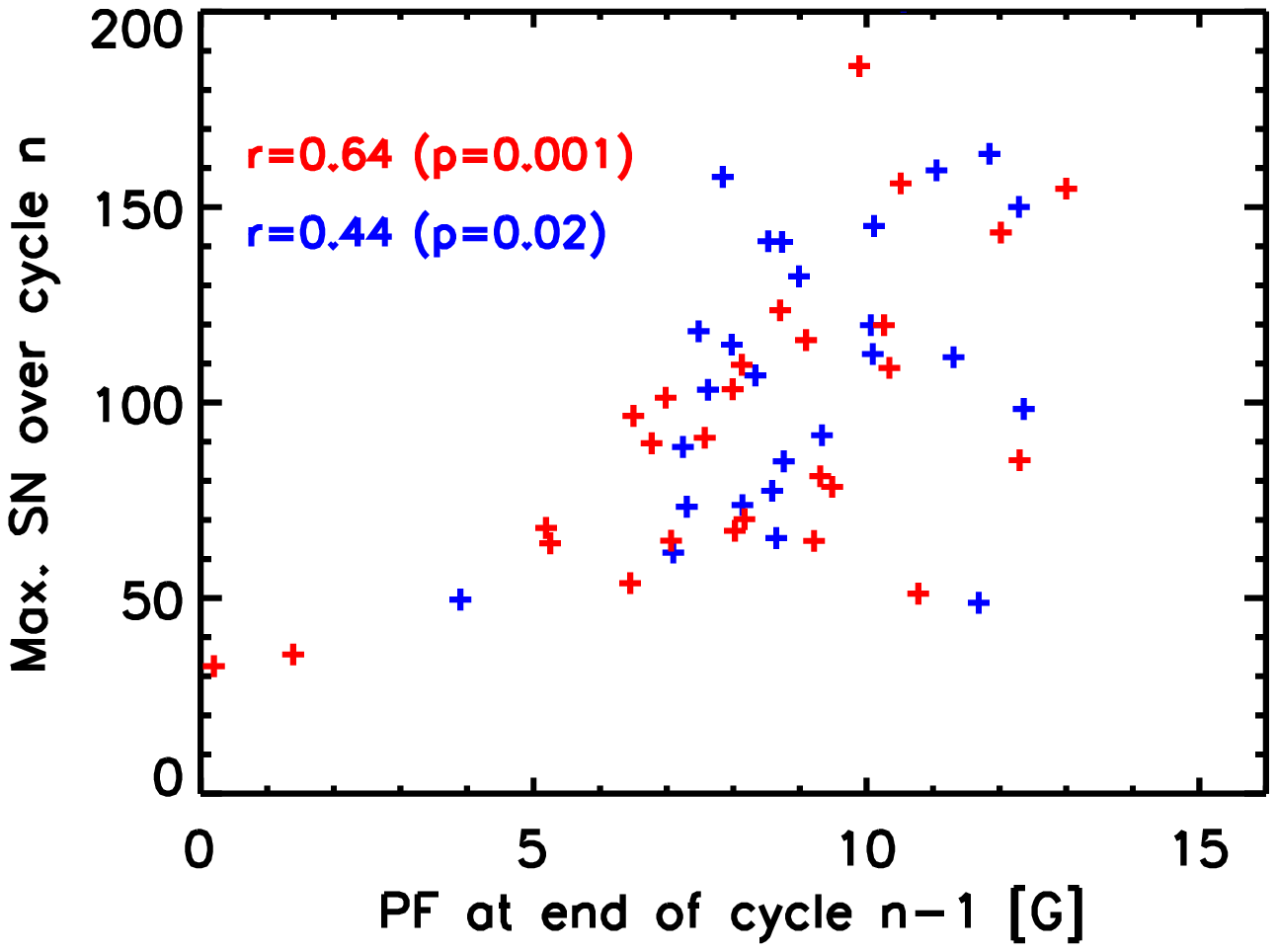}}
\caption{Left: relation between the maximum sunspot numbers of cycle $n$
and the maximum of the polar field at the end of cycle $n$ for the $R_Z$
case (blue symbols) and the $R_G$ case (red symbols). Middle: relation
between maximum sunspot numbers of cycle $n$ and the sum of the maximum
polar fields of cycles $n-1$ and $n$. Right: relation between the
maximum polar field of cycle $n$ and the maximum sunspot number of cycle
$n+1$. Correlation coefficients, $r$, and significance levels, $p$ are
indicated.}
\label{fig:corr_sn_pf}       % Give a unique label
\end{figure*}

The relation between the maximum polar field at the end of a cycle and
the cycle strength (highest sunspot number during the cycle) is shown in
the left panel of Figure \ref{fig:corr_sn_pf}.  The middle panel gives
the relation between the cycle strength and the sum of the maximum polar
field of the preceding and the same cycle. This represents the change in
the polar field from one cycle to the next. The correlation is expected
to be stronger in this case, because the tilted BMRs first reverse the
polar field of the old cycle and then built up that of the new cycle. The
right panel of Figure \ref{fig:corr_sn_pf} shows the relation between
the maximum polar field at the end of a cycle and the strength of the
next cycle. All correlations are significant at the $p\le 0.05$ level.
From the results described in Section 3, we expect that for $\eta_R=0$
the correlation between polar field and strength of the next cycle would
be even stronger. Such a correlation is potentially relevant in
connection with flux transport dynamo models
\citep[e.g.,][]{Chatterjee04,Dikpati04}.

\begin{figure}
\resizebox{\hsize}{!}{\includegraphics{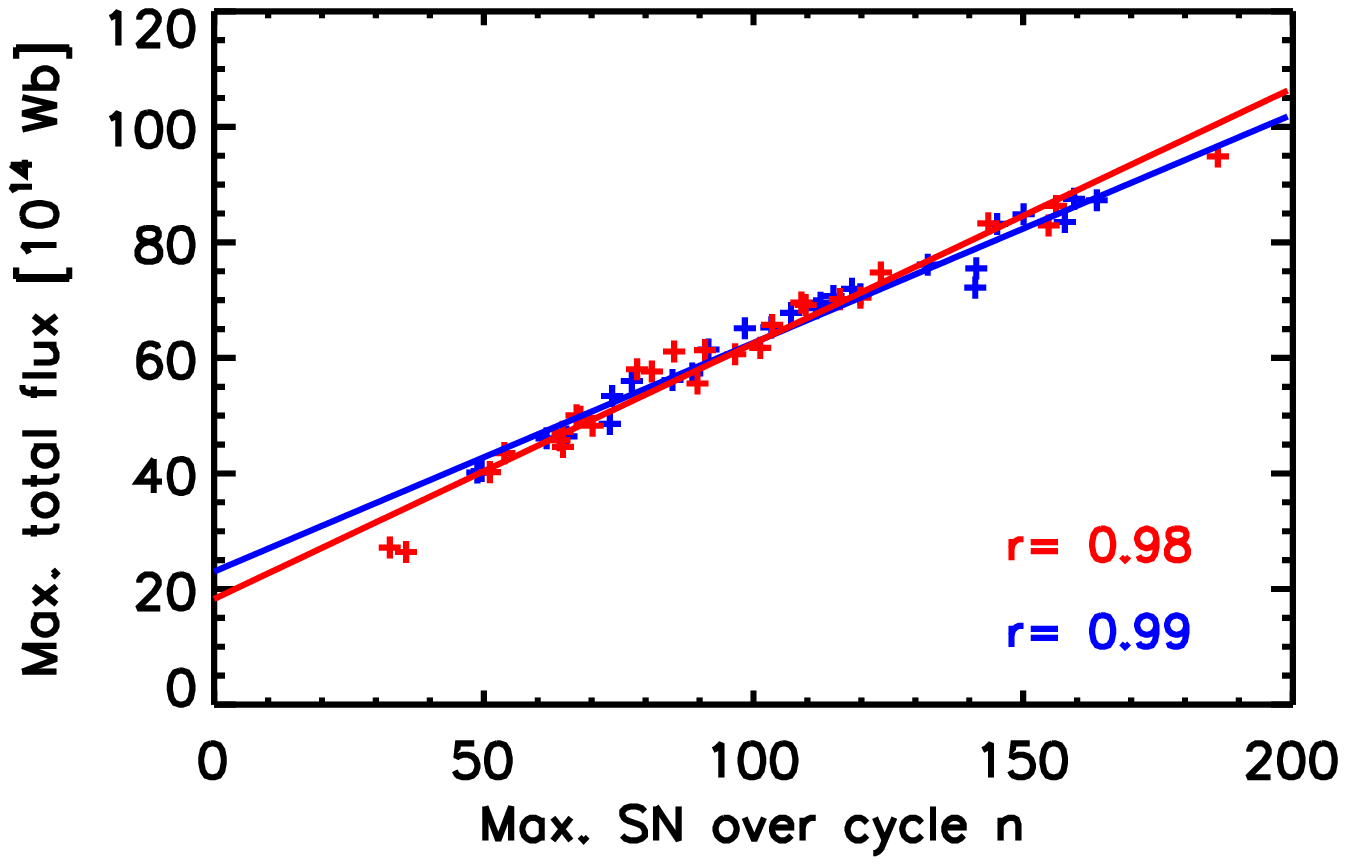}}
\resizebox{\hsize}{!}{\includegraphics{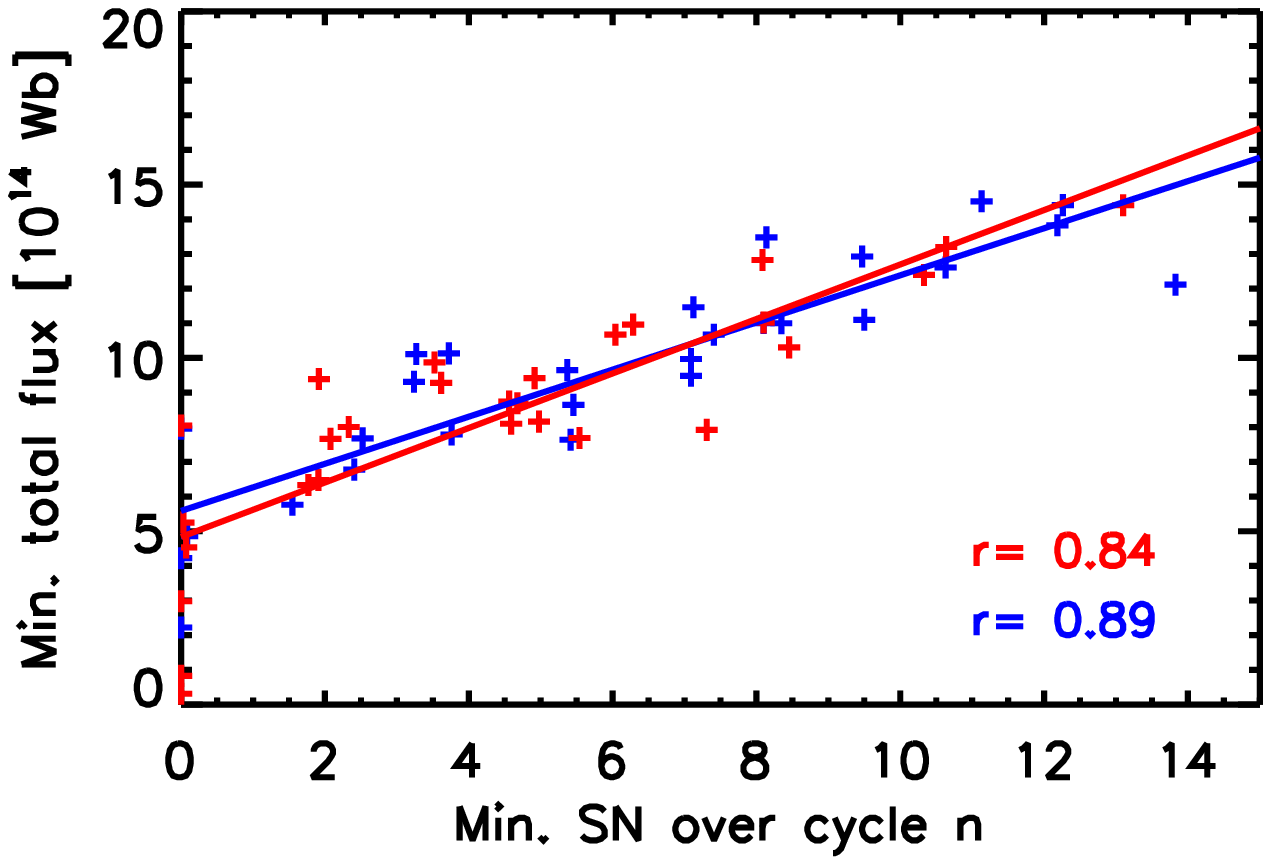}}
\caption{Relation between sunspot number and total surface flux during
cycle maxima (upper panel) and during cycle minima (lower panel). Red symbols refer
to the $R_G$ case and blue symbols to the $R_Z$ case.}
\label{fig:corr_sn_tf}       % Give a unique label
\end{figure}

Figure \ref{fig:corr_sn_tf} shows the relationship between the
total unsigned flux and the sunspot number. During activity maxima, many
BMRs emerge and the polar fields are weak.  
Hence there is a strong correlation between the sunspot number and the
total unsigned flux. During the minima fewer sunspots emerge, and
the total surface flux has a stronger contribution from the polar region.
This leads to a weaker correlation between the total surface field
and the sunspot numbers during solar minima. On longer timescales,
the total flux is roughly proportional to the sunspot
number. 

\begin{figure}
\resizebox{\hsize}{!}{\includegraphics{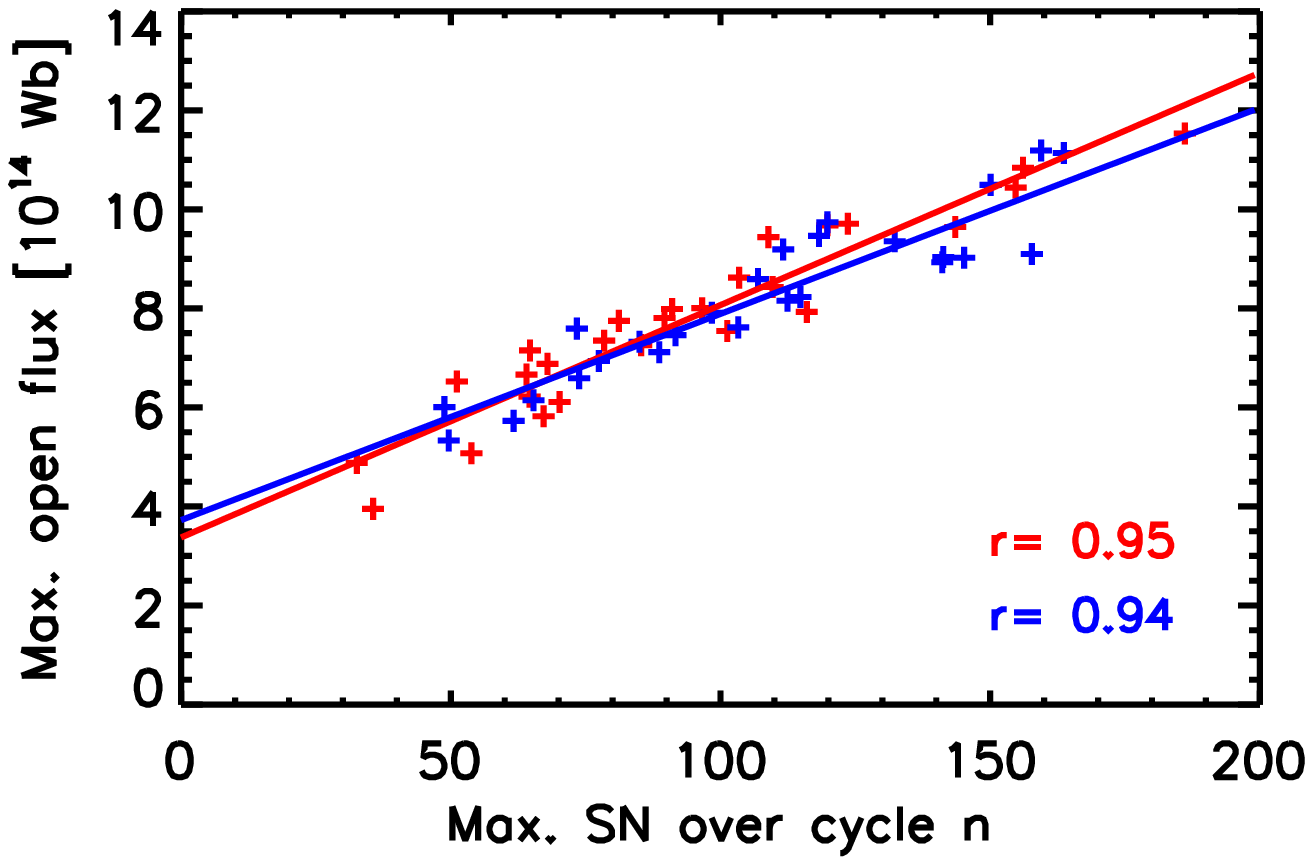}}
\resizebox{\hsize}{!}{\includegraphics{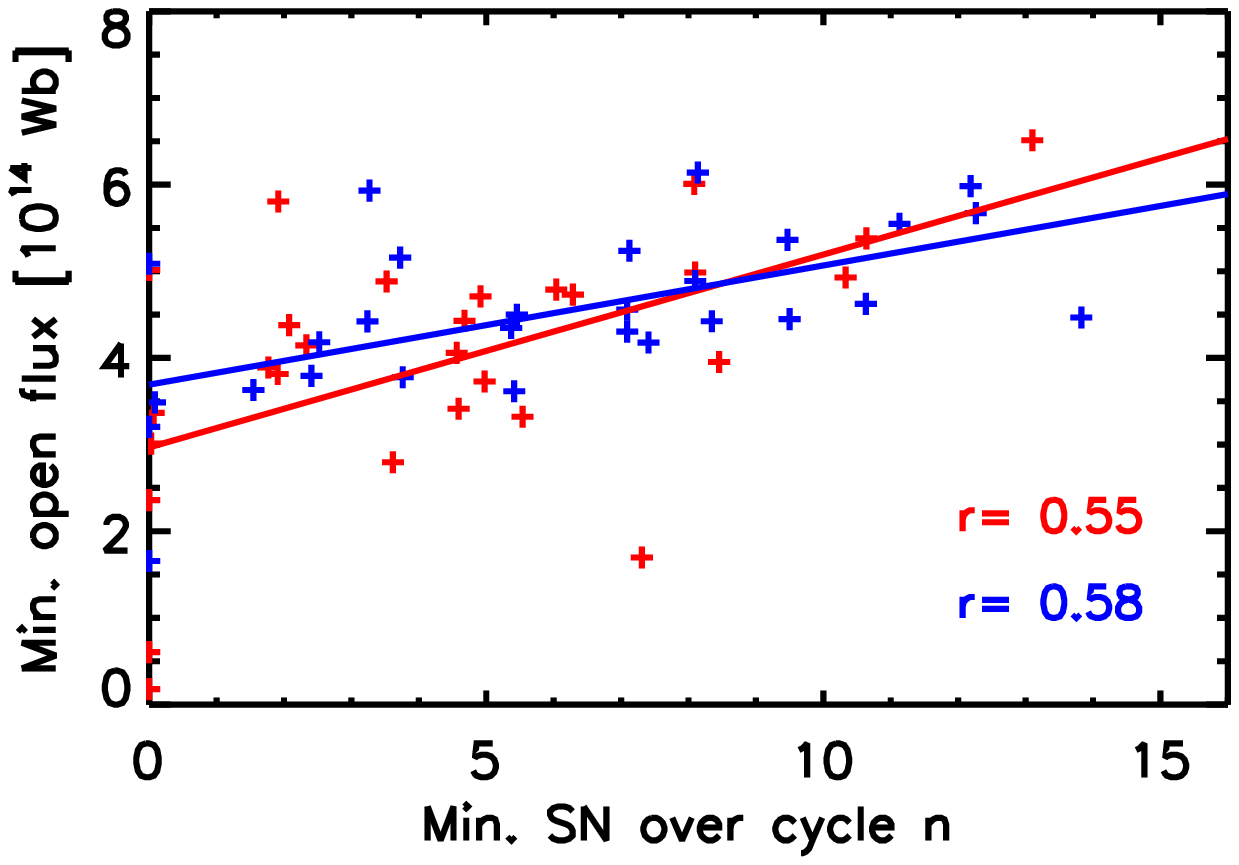}}
\caption{Top: relation between maximum sunspot number and the maximum
open flux during a cycle. Bottom: relation between minimum sunspot
number and the minimum open flux during a cycle.}
\label{fig:corr_sn_of}       % Give a unique label
\end{figure}

Since the contribution to the field strength of the multipole of order
$l$ falls off with radius as $r^{-(l+2)}$, the lowest-order multipoles
dominate the amplitude of the open heliospheric flux \citep{Wang02}. The
emergence of any given BMR may increase or decrease the open flux,
depending on whether its dipole moment vector is oriented so as to
reinforce or reduce that of the pre-existing field \citep{Wang00}. The
left panel of Figure \ref{fig:corr_sn_of} shows that the maximum value of
open flux during a cycle and the maximum sunspot number are strongly
correlated. This results from the large number of BMRs emerging during
the solar maximum period together with activity nesting, which
generate a strong equatorial dipole field. On the other hand, the
correlation is much weaker between the minimum value of open flux over a
cycle and the minimum sunspot number. This is because the open flux
during minimum phases is dominated by the axial dipole field
corresponding to the polar field that has accumulated over the cycle.
Note that the maxima of the open flux usually occur 1--2 yrs after
sunspot maximum \citep[cf.][]{Wang02} while the minima of the open flux
are almost in phase with the sunspot numbers.

\subsection{Comparison of open flux with other reconstructions}
\label{sec:compare_field} 
\begin{figure*}
\resizebox{\hsize}{!}{\includegraphics[angle=90]{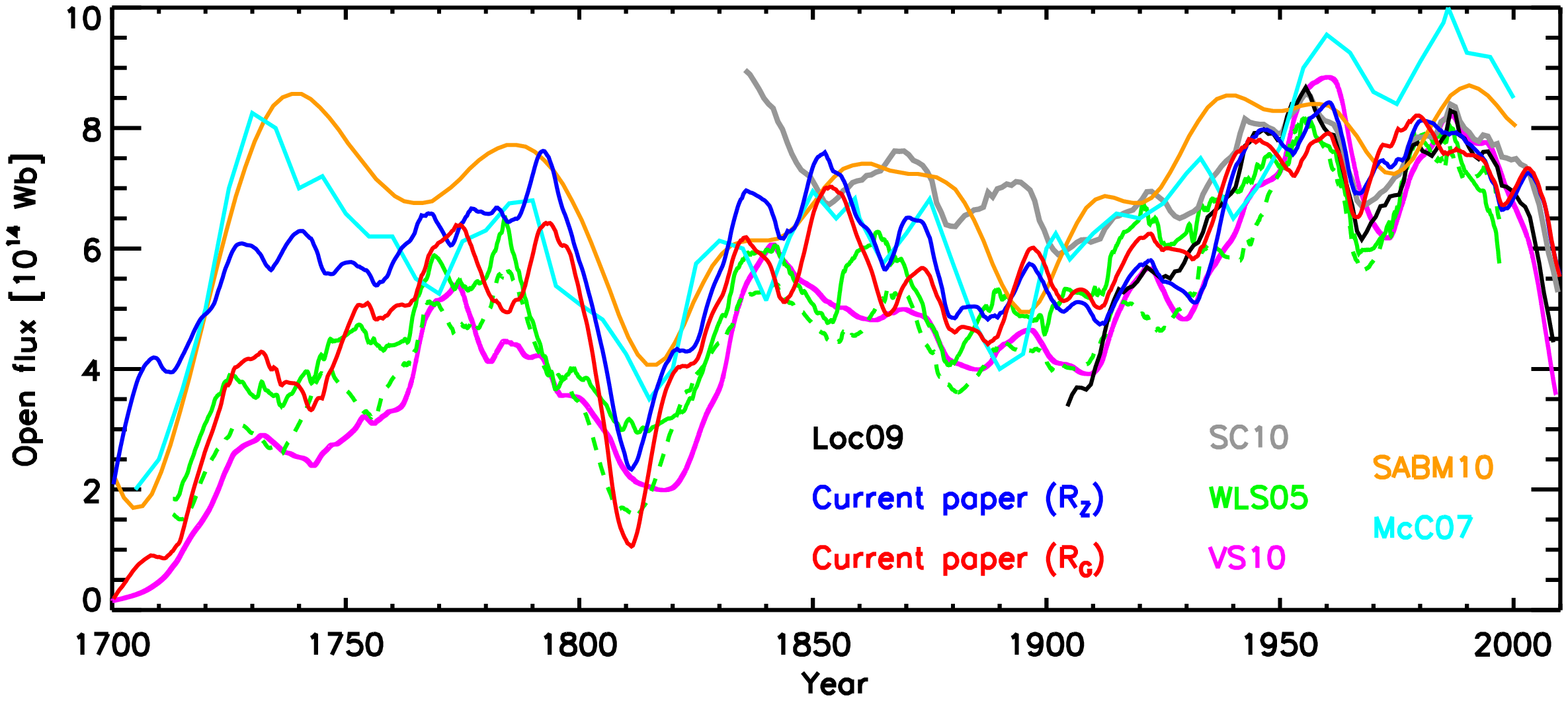}}
\caption{Comparison of the reconstructed open flux with other
reconstructions. Loc09: see \citet{Lockwood03} and \citet{Lockwood09b}. WLS05: \citet{Wang05};
solid and dashed green are their models S1 and S2, respectively. SC10:
\cite{Svalgaard10}. A factor 0.4 used to convert their data into open
flux.  VS10: \citet{Vieira10}. SABM10: \citet[][PCA
composite]{Steinhilber10}. McC07: \citet{McCracken07}. 
All data are 11-yr running averages, except SABM10 which
is a 25-yr running average.}
\label{fig:com_ofs}       % Give a unique label
\end{figure*}

Figure \ref{fig:com_ofs} shows the comparison of our 11-yr running
average of the reconstructed open flux since 1700 with results based on
geomagnetic $aa$ index by \citet[][Loc09]{Lockwood09b} and
\citet[SC10]{Svalgaard10}, from the cosmogenic isotope $^{10}$Be data by
\citet[McC07]{McCracken07} and \citet[SABM10]{Steinhilber10}, and from
other models based on sunspot numbers by \citet[VS10]{Vieira10} and
\citet[WLS05]{Wang05}.  Our reconstructed open flux refers to the
unsigned radial component of the magnetic field at the orbit of
Earth. Since the values given in SC10 correspond to the heliospheric
vector magnetic field amplitude, a factor 0.4 has been applied.  This
factor is close to the value used by \citet{Svalgaard06} to convert the
magnitude of the vector field to the radial component.

All reconstructions show a similar behavior in the second half of the
time period covered. The values given by SABM10 are systematically
higher than the others by approximately 10\%.  The minimum open flux
over the period studied is about 1--2$\times10^{14}$ Wb and the maximum
value is about 8$\times10^{14}$ Wb.

Before $\sim1800$, there is a distinct difference between the results
based on $^{10}$Be data (McC07, SABM10) and the other methods. The
values inferred from $^{10}$Be are particularly high from 1720 to 1750,
when the $R_G$ data indicate rather low solar activity.  Although the
$R_Z$ values are somewhat higher than $R_G$ during this time period, our
reconstructed open flux is still about 30\% weaker than that of McC07
and SABM10.

\subsection{A `lost cycle'?}
\label{sec:lost_cycle}

Since the years 1790--1794 are poorly covered by sunspot observations, it
has been suggested that the unusually long cycle number 4
(1784.7--1798.3) may actually consist of two shorter cycles, so that a
weak `lost' cycle could be missing in the existing sunspot number
records \citep{Usoskin01}.  There is an ongoing discussion on this topic
\citep{Krivova02, Arlt08, Usoskin09}, so that it seems reasonable to
investigate how adding a `lost' cycle $4^{\prime}$ would affect our
reconstructions.

Using $R_G$ data, we followed \citet{Usoskin01} and changed the cycle
minimum times to 1784.3 (cycle 4), 1793.1 (cycle $4^{\prime}$, the possible
lost cycle) and 1799.8 (cycle 5). All other cycle minima times were left
unchanged (for values, see Table~1 of Paper I). We then generated new
semi-synthetic sunspot group records according to the method of Paper I.
In accordance with Hale's polarity laws, the magnetic polarity
orientation of the sunspot groups in cycle $4^{\prime}$ was taken to be the
same as that of the original cycle 4 and the orientations for all
cycles before cycle $4^{\prime}$ were reversed.

\begin{figure}
\resizebox{\hsize}{!}{\includegraphics{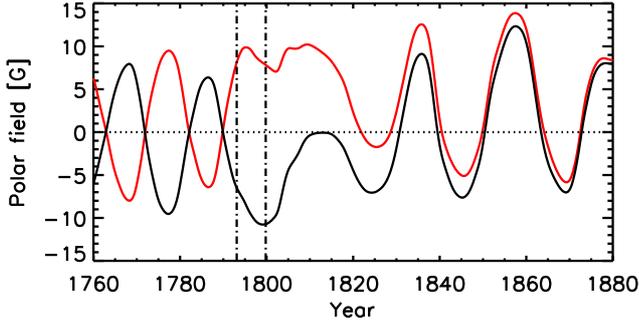}}
\caption{Time evolution of the reconstructed (southern) polar field
during 1760--1880 after adding the `lost' cycle indicated by the two vertical
lines between 1793.1 and 1799.8 (red curve) and without the additional
cycle (standard case, black curve).  Both reconstructions are based on
$R_G$.}
\label{fig:field_lost_bpol}       % Give a unique label
\end{figure}

Figure \ref{fig:field_lost_bpol} shows the comparison of the time
evolution of the polar field between the cases with and without cycle
$4^{\prime}$.  The polar field generated during cycle 4 is only slightly
diminished by the short and weak cycle $4^{\prime}$ and no reversal takes
place. The following cycle 5 has the same polarities as cycle 4 and
thus strengthens the polar field even further. Thus the addition of the
extra cycle leads to a strong polar field being maintained throughout
the whole period between about 1790 and 1810. This is followed by a
period of weak polar field between 1820 and 1830. In contrast to these
results, the standard case without extra cycle predicts weak polar field
between 1805 and 1820. 

\begin{figure}
\resizebox{\hsize}{!}{\includegraphics{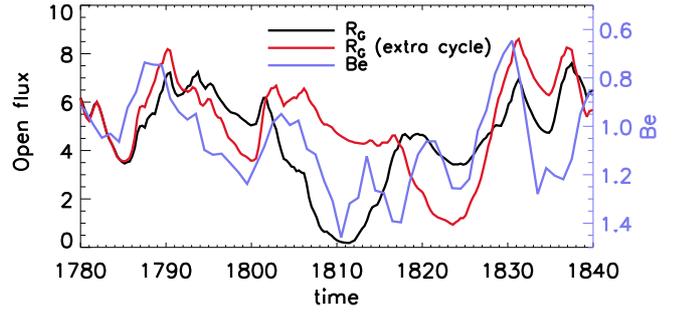}}
\caption{Time evolution of the reconstructed open heliospheric flux
(11-yrs running average) during 1780--1840 with the `lost' cycle (red
curve) and without (black curve), in comparison to the measured
$^{10}$Be concentration in ice cores \citep{Beer90}. Note the
inverted scale for the $^{10}$Be data.  }
\label{fig:field_lost_open}       % Give a unique label
\end{figure}

Comparing the reconstructed open heliospheric flux with the measured
$^{10}$Be concentration in ice cores should, at least in qualitative
terms, give an indication which model is to be preferred.  Figure
\ref{fig:field_lost_open} shows the evolution of 11-yr running average
of the reconstructed open flux in both cases together with the $^{10}$Be
data of \citet{Beer90}. The
introduction of the extra cycle shifts the minimum of the open flux from
about 1810 to about 1825, which is clearly inconsistent with the
$^{10}$Be data showing maxima (corresponding to low open flux) around
1810 and 1817. The same inconsistency arises also when comparing with
the other reconstructions of the open flux shown in
Figure~\ref{fig:com_ofs}.

The strong unreversed polar field and the delay of the open flux minimum
at the beginning of the 19th century by $\sim15$ years, which result
from the introduction of extra cycle $4^{\prime}$, appear to be highly
anomalous and seem to be ruled out by comparison with the $^{10}$Be
record. Our reconstructions therefore argue against the existence of a
`lost' cycle $4^{\prime}$.

\section{Possible sources of error in the reconstruction}
\label{sec:err_ana} 
Our reconstructions are based on the datasets of the sunspot numbers
$R_Z$ and $R_G$. Both of these have errors, partly due to sparse
observations during the early period \citep{Vaquero07} and partly due to
subjective definitions of what to include by the different observers
over the centuries \citep{Svalgaard10}.  Errors in sunspot numbers
obviously affect our reconstruction of the magnetic field. In the
following, we analyze the effect of possible errors in the sunspot
numbers on the reconstructed magnetic field by considering a
weak cycle (cycle 6, 1810.6--1823.3) and a strong cycle (cycle 19,
1954.3--1964.9) as extreme examples. We restrict this analysis to the
reconstruction based on $R_Z$, noting that the results for the $R_G$
case are very similar.

\begin{figure*}
\centering
\resizebox{\hsize}{!}{\includegraphics{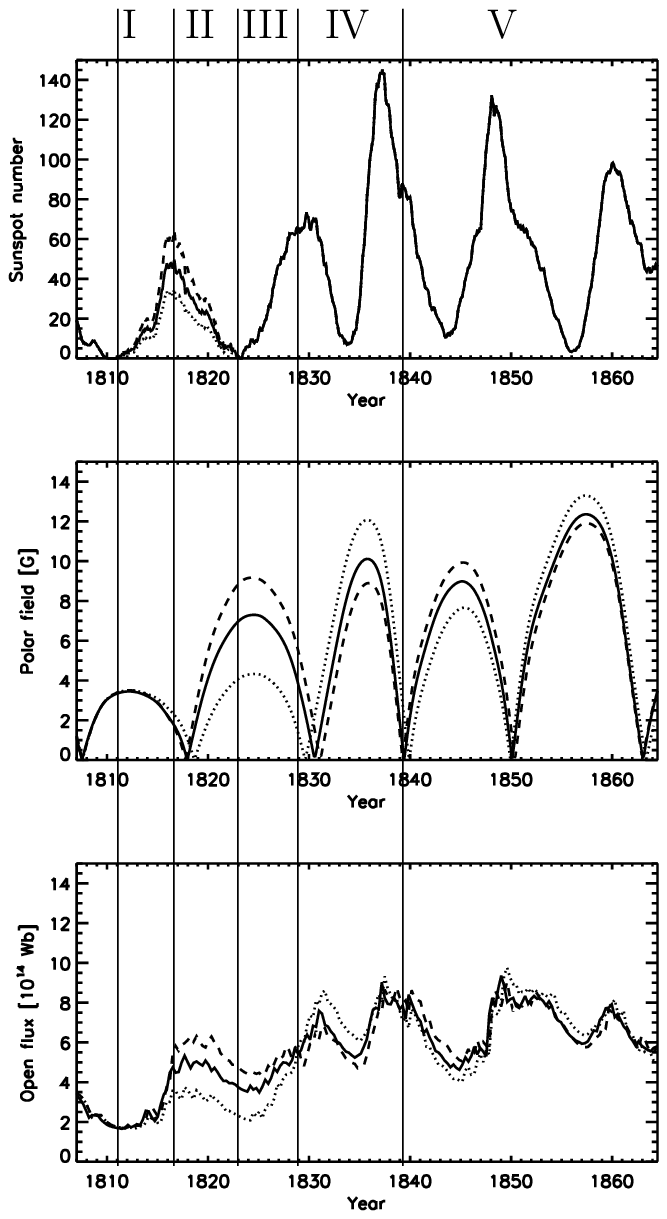}\includegraphics{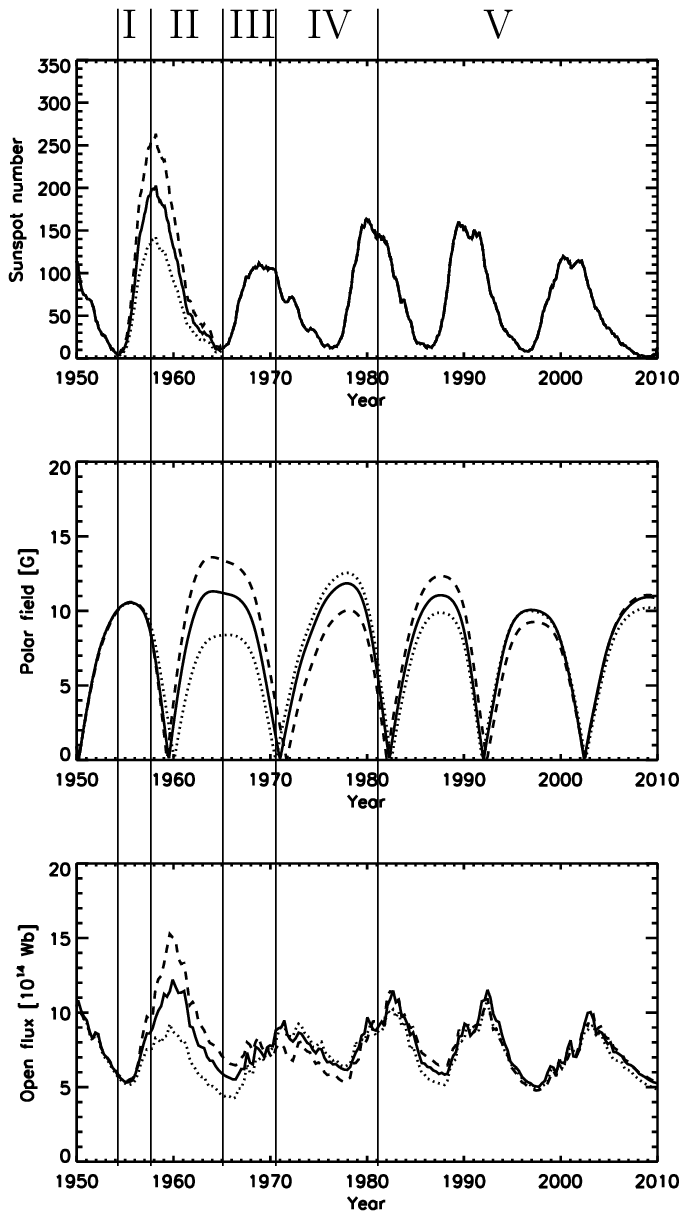}}
\caption{Effect of varying the sunspot number (upper panels)
by an increase of 30\% (dashed curves) and by a decrease of 30\% (dotted
curves) on the evolution of the polar field (middle panels, average
absolute value of the northern and southern poles) and the open flux
(lower panels). The weak cycle 6 (1810.6--1823.3, left panels) and the
strong cycle 19 (1954.3--1964.9, right panels) of the $R_Z$ case are
considered. The Roman numbers denote 5 time intervals which are affected
by the variation of the sunspot numbers in different ways (see text).}
\label{fig:err1}       % Give a unique label
\end{figure*}

The effects of a change of the sunspot number data on the polar field
and the open flux in cycle 6 (left panels) and cycle 19 (right panels)
are displayed in Figure \ref{fig:err1}. The upper panels illustrate a
reduction of the sunspot number by 30\% (dotted curves) and an increase
by 30\% (dashed curves), respectively, of the original values. The solid
curves represent the unchanged data. The previous 5 yr before the cycle
starts and the subsequent 45 yr after the cycle ends are shown as well.

To simplify the discussion, we denote cycle 6 and cycle 19 as cycle $n$,
cycles 7 and 20 as $n+1$, and so on. The effect of the variation of
sunspot number of cycle $n$ on the following cycles has a periodicity of
two cycles and an amplitude which decays with an $e$-folding time of
$\sim20$ years. To reveal the effects in more detail, we have broken the
time period into 5 intervals. Part \textrm{I} corresponds to the rising
phase of cycle $n$ (from the start to the maximum). The polar field
(shown in middle panels of Figure \ref{fig:err1}) is only weakly
affected during this interval, during which the emerging BMRs reverse
the polar field of the previous cycle. Hence higher sunspot numbers
cause the polar field to become weaker and to reverse earlier.  The
variation of the open flux (lower panels of Figure \ref{fig:err1})
during this time period is dominated by the variation of equatorial
dipole field which is directly connected to the sunspot number. Due to
the competing effects of the axial dipole field (polar field), the
relative variation of the open flux is less than that of the sunspot
number.

Period \textrm{II} corresponds to the decaying phases of cycle $n$ (from
the maximum to the end of cycle). Part \textrm{III} is the rising phase
of cycle $n+1$. During these two periods, the polar field and the open
flux are strongly affected. More sunspots produce a stronger polar field
and vice versa. The relative variation of the open flux is similar to
the relative change of the sunspot number during period \textrm{II}, but
the effect is weaker for part \textrm{III} since during the rise to the
next maximum the equatorial dipole moment of cycle $n+1$, which is
unchanged, begins to dominate.

Period \textrm{IV} mainly includes the decaying phases of cycle $n+1$
and rising phases of cycle $n+2$. The polar field is still significantly
affected but less so than during the periods \textrm{II} and
\textrm{III}, owing to the effect of radial diffusion. More sunspots in
cycle $n$ produce a stronger polar field during this time.  In period
\textrm{V}, the effect of a variation of the sunspot number in cycle $n$
becomes less and less important owing to the decay caused by the radial
diffusion.

\begin{figure}
\centering
\resizebox{\hsize}{!}{\includegraphics[angle=90]{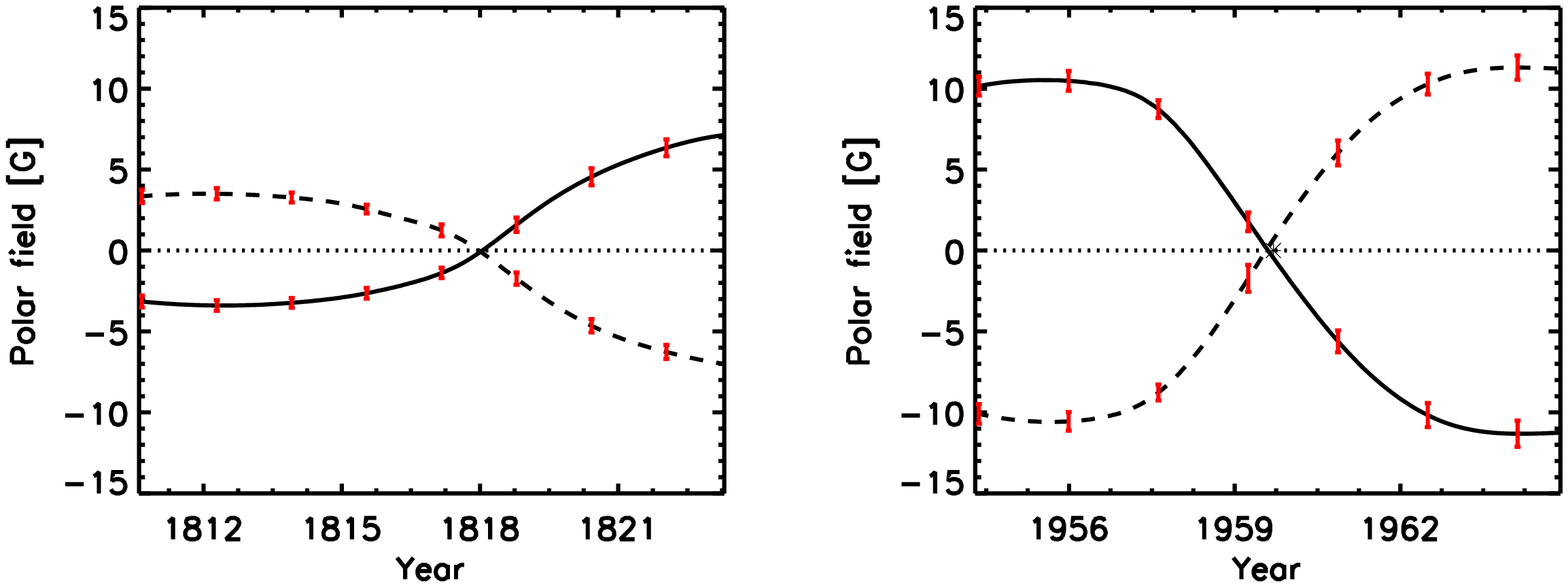}}
\resizebox{\hsize}{!}{\includegraphics[angle=90]{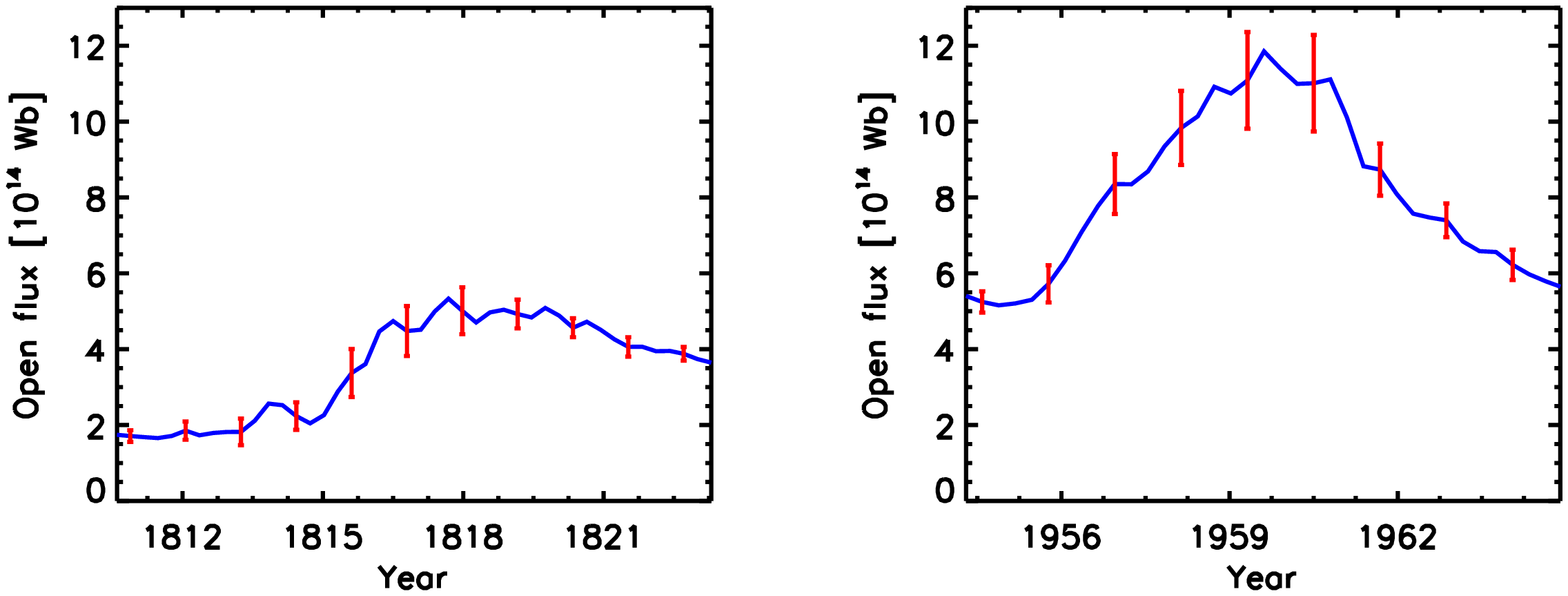}}
\caption{The effect of the random component in generating the sources
for the SFTM.  The reconstructed polar field (upper panels) and open flux
(lower panels) are shown for cycle 6 (left) and cycle 19 (right). The
red vertical bars indicate the standard deviations corresponding to a set
of twenty synthetic sunspot group records with different random
numbers.}
\label{fig:err2}       % Give a unique label
\end{figure}

In addition, the derivation of the sources for the SFTM depends on the
spatial and temporal properties of BMR emergences. Because we derive the
source in terms of correlations, the model has random components. Figure
\ref{fig:err2} shows the mean values and standard deviations arising
from this randomness, derived using 20 independent realizations of
the semi-synthetic sunspot group records.  Again the results for a weak
cycle (cycle 6) and a strong cycle (cycle 19) are shown.  The standard
deviations are small for the polar field, but more significant for the
open flux during the maximum periods and reach up to 13\% of the open
flux. The average standard deviations of the polar field and the open
flux over the whole simulation time series are about 10\% of the 
corresponding mean values.

\section{Conclusion}

We have provided a physical reconstruction of the large-scale solar
magnetic field and the open heliospheric flux since 1700 with a surface flux
transport model with sources based on sunspot number data and on the
statistical properties of the sunspot groups in the RGO
photoheliographic results. The model has been validated
through comparison with reconstructions based on the actual sunspot
group record and with directly measured or observationally inferred
quantities. 

Our source term, $S$, is based on the semi-synthetic sunspot group record 
of Paper I. It in turn is based upon correlations between the cycle amplitudes and 
cycle phase, and sunspot group areas, emergence latitudes and tilt angles.  
Hence while the surface evolution described by the  SFTM model is linear, 
our model of the source term introduces nonlinearities into the reconstruction. 
These nonlinearities partly explain why 
the reconstruction for the period from 1874 to 1976 has polar field reversals
each cycle without the need to invoke variations in the meridional flow
\citep{Wang02} or extra terms which cause the field to decay 
\citep{Schrijver02}. This was also found in CJSS10.
The sunspot numbers are less reliable prior to 1874, and we needed to introduce 
a slow decay of the field to produce polar fields which reverse each cycle. 

The reconstructions considerably extend
the basis for correlations studies, such as the relation between 
the polar field amplitude during activity minima and the strengths
of the preceding and subsequent cycles, with implications for dynamo
models. Introducing a possibly `lost' cycle at the end of the 18th
century leads to a shift of the open flux minimum during the Dalton
minimum which is incompatible with the $^{10}$Be record.  

\begin{acknowledgements}
We are grateful to Y.-M. Wang, L. E. A. Vieira,
M. Lockwood, and F. Steinhilber for kindly providing their reconstructed open flux data.
\end{acknowledgements}

\bibliographystyle{aa}
\bibliography{16168_bib}

\end{document}